\documentclass[sigconf]{acmart}
\usepackage{multirow}
\usepackage{balance}
\usepackage{caption}
\usepackage{subcaption}
\usepackage{algorithm}
\usepackage{algorithmic}

\newcommand{\ie}{\textit{i.e.}}
\newcommand{\eg}{\textit{e.g.}}
\newcommand{\ttt}{\texttt}

\AtBeginDocument{%
  \providecommand\BibTeX{{%
    \normalfont B\kern-0.5em{\scshape i\kern-0.25em b}\kern-0.8em\TeX}}}

\copyrightyear{2022}
\acmYear{2022}
\setcopyright{acmlicensed}
\acmConference[CIKM '22] {Proceedings of the 31st ACM International Conference on Information and Knowledge Management}{October 17--21, 2022}{Atlanta, GA, USA.}
\acmBooktitle{Proceedings of the 31st ACM International Conference on Information and Knowledge Management (CIKM '22), October 17--21, 2022, Atlanta, GA, USA}
\acmPrice{15.00}
\acmISBN{978-1-4503-9236-5/22/10}
\acmDOI{10.1145/3511808.3557328}
\begin{document}
\title[From Easy to Hard: A Dual Curriculum Learning Framework for Context-Aware Document Ranking]{From Easy to Hard: A Dual Curriculum Learning Framework \\ for Context-Aware Document Ranking}

\author{Yutao Zhu}
\author{Jian-Yun Nie}
\affiliation{University of Montreal \city{Montreal} \state{Quebec} \country{Canada}}
\email{yutaozhu94@gmail.com}
\email{nie@iro.umontreal.ca}
\author{Yixuan Su}
\affiliation{Language Technology Lab, University of Cambridge \city{Cambridge} \country{United Kingdom}}
\email{ys484@cam.ac.uk}
\author{Haonan Chen}
\affiliation{Gaoling School of Artificial Intelligence, Renmin University of China \state{Beijing} \country{China}}
\email{hnchen@ruc.edu.cn}
\author{Xinyu Zhang}
\affiliation{Huawei Poisson Lab \city{Hangzhou} \state{Zhejiang} \country{China}}
\email{zhangxinyu35@huawei.com}
\author{Zhicheng Dou}
\affiliation{Gaoling School of Artificial Intelligence, Renmin University of China \state{Beijing} \country{China}}
\email{dou@ruc.edu.cn}


\renewcommand{\shortauthors}{Yutao Zhu et al.}
\renewcommand{\authors}{Yutao Zhu, Jian-Yun Nie, Yixuan Su, Haonan Chen, Xinyu Zhang, and Zhicheng Dou}



\begin{abstract}
Contextual information in search sessions is important for capturing users' search intents. Various approaches have been proposed to model user behavior sequences to improve document ranking in a session. Typically, training samples of (search context, document) pairs are sampled randomly in each training epoch. In reality, the difficulty to understand user's search intent and to judge document's relevance varies greatly from one search context to another. Mixing up training samples of  different difficulties may confuse the model's optimization process. In this work, we propose a curriculum learning framework for context-aware document ranking, in which the ranking model learns matching signals between the search context and the candidate document in an easy-to-hard manner. In so doing, we aim to guide the model gradually toward a global optimum. To leverage both positive and negative examples, two curricula are designed. Experiments on two real query log datasets show that our proposed framework can improve the performance of several existing methods significantly, demonstrating the effectiveness of curriculum learning for context-aware document ranking.
\end{abstract}





\maketitle

\section{Introduction}
Users' search behaviors have evolved from one-shot queries to multiple interactions with search engines~\cite{DBLP:conf/sigir/AgichteinWDB12}. To fulfill a complex information retrieval task, users may issue a series of queries, examine and interact with some results.
Many studies have shown that  historical user behavior or search activities can be leveraged to improve document ranking, especially when users' search intent is ambiguous~\cite{DBLP:conf/cikm/JonesK08,DBLP:conf/sigir/BennettWCDBBC12,DBLP:conf/cikm/GeDJNW18,CARS,DBLP:conf/cikm/ZhouD0W21,HQCN}. 

Previous studies have exploited user search behaviors for understanding user intent and improving document ranking within a session~\cite{DBLP:conf/sigir/BennettWCDBBC12,DBLP:conf/sigir/CarteretteCHKS16,DBLP:conf/cikm/SordoniBVLSN15}. Earlier research explored query expansion and learning-to-rank techniques~\cite{DBLP:conf/sigir/BennettWCDBBC12,DBLP:conf/sigir/CarteretteCHKS16,DBLP:conf/ictir/GyselKR16,DBLP:conf/sigir/ShenTZ05}. More recently, many neural architectures have been developed to model user behavior sequences and capture user search intent~\cite{DBLP:conf/cikm/SordoniBVLSN15,CARS,MNSRF,HBA,COCA}. 
For example, a hierarchical RNN with attention mechanism was used to model the historical queries and the corresponding clicked documents, leading to better 
document ranking~\cite{DBLP:conf/cikm/SordoniBVLSN15,MNSRF}. Researchers also discovered that learning query suggestion as a supplementary task is also beneficial for document ranking~\cite{CARS}. Recently, pre-trained language models have also been used to capture search intent from user behavior sequences~\cite{HBA,COCA}.

\begin{figure*}[t!]
    \centering
    \includegraphics[width=.75\linewidth]{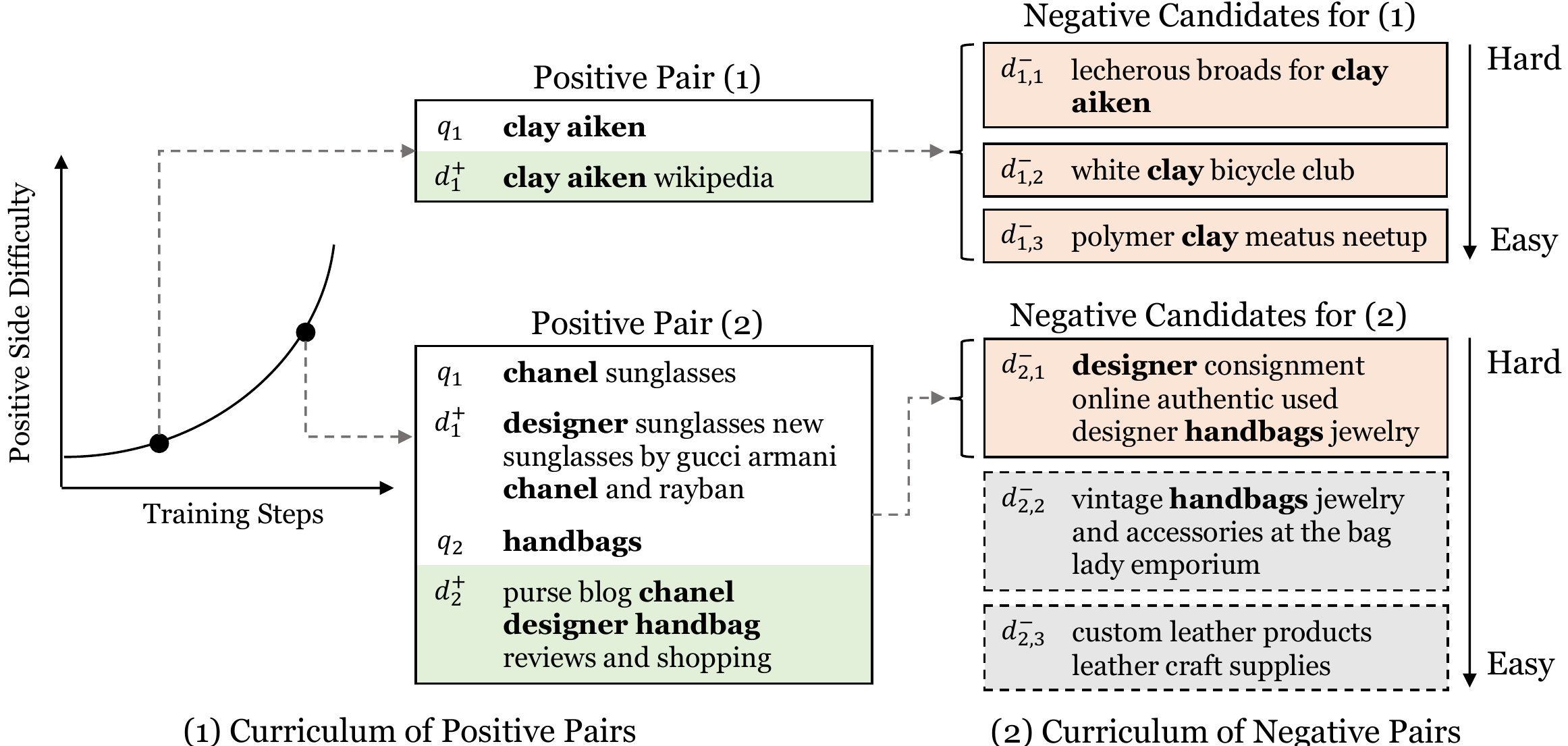}
    \caption{Illustration of \ttt{DCL}'s utilization of samples. The left side shows  positive pairs of different difficulties. In the early training steps, the model can only learn from easy pairs (\eg, the first pair). Then, as the training progresses, harder pairs (\eg, the second pair) are added. The right side is the curriculum of negative pairs. All negative candidates ($d_{1,1}^-$, $d_{1,2}^-$, and $d_{1,3}^-$) of easy cases are used for training in the early steps, while only hard negatives ($d_{2,1}^-$) are used in later steps.}
    \label{fig:method}
    \vspace{-10px}
\end{figure*}

Although the approaches proposed are different, they all rely on the information extracted from previous search logs, for example, (search context, document) pairs that are considered as positive or negative samples. The search context usually aggregates the current query and the previous queries (and interacted documents in some cases). All positive pairs are assumed to be samples of equal importance that reflect relevance, and are put together in the same pool for sampling a training batch. The same for negative pairs that reflect irrelevance. While it is true that positive and negative examples are generally useful for training a good ranking model, it is not true that they are equally useful at different training stages. In some cases, the relevance (or irrelevance) relation in a pair of (search context, document) is obvious, while in other cases, it is more subtle. 
Let us illustrate this by some examples. In the left part of Figure~\ref{fig:method}, we show two search contexts (sessions) containing queries and the documents clicked by the user. Let us try to understand the search intent in each of the search context. In the first case, the relevance relation between the search context formed by $q_1$ and the clicked document $d_1^+$ is clear because both are about the singer ``Clay Aiken''. In comparison, in the second case, the relation between the search context formed by $[q_1, d_1^+, q_2]$ and the clicked document $d_2^+$ is more difficult to capture. The underlying intent is ``Chanel's designer handbags'', which can only be understood with the help of the historical query $q_1$ and the clicked document $d_1^+$. If a human is asked to learn from these two samples, the relevance signal in the first one is much easier to capture than in the second. For a training process, the situation is similar: These pairs represent different levels of \textit{difficulty} for a training process to digest. 
When a human learner is presented with samples of mixed difficulties, he/she can be confused because signals from different samples may appear inconsistent, or they do not have sufficient knowledge to understand difficult samples.
Recent studies in machine learning also showed that learning with a batch of samples of mixed difficulties may disturb the optimization, especially when the network is deep~\cite{DBLP:conf/icml/BengioLCW09}. In this case, curriculum learning, \ie, learning from easy samples before hard samples, becomes particularly useful. 

A similar problem occurs when negative samples are considered. As can be seen in the right side of Figure~\ref{fig:method}, for the second search context, the negative candidate $d_{2,1}^-$ contains information about ``designer handbags''. Compared with $d_{2,3}^-$, we can see that $d_{2,1}^-$ is much harder than $d_{2,3}^-$ to be recognized as a negative sample. Therefore, it is also desirable to learn from easy negative samples before hard ones. 
If we compare $d_{2,1}^-$ with the positive document $d_2^+$, we can learn that ``chanel'' is an important term reflecting the user's real intent. Such comparison/contrast  is also common and critical for human learning, especially for discriminating similar concepts~\cite{DBLP:journals/cogsci/Chin-ParkerC17}. 

Motivated by the observations above, we propose a novel training framework that takes into account the levels of  difficulty in the training samples. Our framework is inspired by both curriculum learning~\cite{DBLP:conf/icml/HacohenW19,DBLP:journals/corr/abs-1912-08555} and contrastive learning~\cite{SimCLR,SimCSE,DBLP:journals/corr/abs-2202-06417,DBLP:conf/naacl/Su0MLSSC22}. Curriculum learning simulates the human recognition process, \ie, learning with easier samples first and more difficult samples later. It has achieved great performance on various tasks, such as image classification~\cite{DBLP:journals/tip/GongTMLKY16,DBLP:conf/icml/HacohenW19}, natural language understanding~\cite{DBLP:conf/acl/XuZMWXZ20}, and ad-hoc retrieval~\cite{DBLP:journals/corr/abs-1912-08555}. Contrastive learning aims at learning representations such that similar samples stay close to each other, while dissimilar ones are far apart. By comparing similar and dissimilar samples, the model can be better optimized to capture their differences.

Concretely, we treat the search context and the candidate document as a pair and optimize the model by contrasting positive pairs (with clicked documents) and negative pairs (with unclicked documents).
We design a dual curriculum learning framework incorporating two complementary curricula for positive and negative pairs, respectively. 
In the curriculum for positive pairs, sampling is restricted to easy pairs in early steps, and then extended gradually to the whole set of samples, so that hard pairs can also be learned. In the curriculum for negative pairs, we do the opposite: sampling from all pairs in early steps, then restricting gradually to hard pairs in late steps.
These strategies are inspired by similar human learning: we select learning (positive) examples from easy to hard, but want to contrast them with all the negative examples at the beginning to have a better idea of the general differences between the positives and negatives. Toward the end of the learning, as the obvious differences have been learned, the easy negatives cannot provide effective learning signals, so we focus on distinguishing hard negative examples (these hard ones are demonstrated to be beneficial for model optimization~\cite{ance,DBLP:conf/sigir/ZhanM0G0M21}). 
Our two curricula intend to follow the same principle during the training process.


The curriculum strategy can be used in any existing approach. We integrate it into three state-of-the-art approaches for context-aware document ranking.
We conduct experiments on two large-scale search log datasets (AOL~\cite{AOL} and Tiangong-ST~\cite{Tiangong}). Experimental results show that our curriculum learning method significantly improves three strong baselines. The consistent performance gains
demonstrate the effectiveness and wide applicability of our approach. Our further experiments show that both positive and negative curricula are beneficial to the ranking effectiveness. 


Our contributions are three-fold:

(1) We propose a novel curriculum learning framework for context-aware document ranking, in which the difficulty of training samples is taken into account. 

(2) We devise two complementary curricula for learning user intent from positive and negative pairs of (search context, candidate document). By learning them in an easy-to-hard manner, the model's performance can be improved gradually. 

(3) Experimental results on two large-scale benchmark datasets show significant improvements. The experiments also confirm the broad applicability, flexibility, and high robustness of our method.

\section{Related Work}
\subsection{Context-Aware Document Ranking}
Context information in sessions has shown to be beneficial in modeling user search intent~\cite{DBLP:conf/cikm/JonesK08,DBLP:conf/sigir/BennettWCDBBC12,DBLP:conf/cikm/GeDJNW18}. Early research focused on characterizing users' search intent by extracting contextual features from their search behaviors~\cite{DBLP:conf/sigir/ShenTZ05,DBLP:conf/cikm/WhiteBD10,DBLP:conf/sigir/XiangJPSCL10}. 
However, because these methods are built on handcrafted rules or manually collected features, they can only be used for a limited number of retrieval tasks.
Researchers also developed predictive models for users' search intent or future behavior~\cite{DBLP:conf/www/CaoJPCL09}, but the learning of complex user-system interactions is limited by the predefined features. 

The recent development of deep neural networks has triggered new approaches to context-aware document ranking. For example, researchers exploited hierarchical RNN-based architectures to model the sequence of historical queries~\cite{DBLP:conf/cikm/SordoniBVLSN15,DBLP:conf/cikm/JiangW18,DBLP:conf/www/WuXS018}. These architectures were further enhanced by attention mechanism to better capture search behaviors~\cite{DBLP:conf/sigir/ChenCCR18}. It is also found that learning query suggestion and document ranking jointly can boost the performance on both tasks~\cite{MNSRF}. Besides, historical clicked documents are also reported to be helpful in predicting user search behaviors~\cite{CARS}. 

Recently, pre-trained language models, such as BERT~\cite{BERT}, have achieved promising results on several NLP and IR tasks~\cite{DBLP:conf/acl/LiuHCG19,DBLP:conf/emnlp/KhashabiMKSTCH20,DBLP:conf/sigir/KhattabZ20,DBLP:conf/wsdm/MaGZFJC21}. Some researchers proposed concatenating all historical queries and candidate documents into a long sequence to compute a sequence representation using BERT, 
based on which the ranking score is determined~\cite{HBA}. Furthermore, contrastive learning has shown to be beneficial for optimizing the BERT encoder in context-aware document ranking~\cite{COCA}. 

Different from the studies above, we focus on improving the model optimization process by curriculum learning rather than designing new architectures or supplementary tasks for context-aware document ranking. 
Our work is orthogonal to the above approaches and can be combined with them. In fact, in most existing approaches, sampling of training data is necessary in the optimization process. This is typically done by random sampling or by selecting hard samples. Our work will show that selecting samples by curriculum from easy to hard can better optimize the existing models.


\subsection{Curriculum Learning for IR}
In the context of human learning, it is common to follow a curriculum that regulates the ordering and content of the education materials~\cite{skinner1958reinforcement,pavlov2010conditioned,krueger2009flexible}. With this strategy, students can leverage previously learned concepts to help them learn new and more difficult ones. Inspired by research in cognitive science~\cite{rohde1999language}, researchers proposed machine learning algorithms based on a curriculum~\cite{elman1993learning,DBLP:conf/icml/BengioLCW09}. The core idea is to train the model using easy samples first and increase the difficulty along the training process. Such a curriculum learning (CL) strategy has achieved great performance on several tasks, such as image classification~\cite{DBLP:journals/tip/GongTMLKY16,DBLP:conf/icml/HacohenW19}, machine translation~\cite{DBLP:journals/corr/abs-1811-00739,DBLP:conf/naacl/PlataniosSNPM19}, dialogue generation~\cite{DBLP:conf/acl/Su0ZLBC0C020}, and natural language understanding~\cite{DBLP:conf/acl/XuZMWXZ20}.

In the area of IR, CL research is still in its early stage. The first attempt applied CL to learning-to-rank (LTR)~\cite{DBLP:conf/cikm/FerroLM018}, however, without much success.
Later, researchers found that manually collected features in LTR can be a source of noise, and the CL strategy is more suitable for neural ranking models~\cite{DBLP:journals/corr/abs-1912-08555}. More recently, several heuristics have been proposed to determine the difficulty of different answers, based on which a CL-based method is used. This led to improved performance in answer ranking~\cite{DBLP:conf/sigir/MacAvaneyN0TGF20a}.


All existing CL-based methods for IR tasks are designed for organizing positive samples, but the influence of negative samples is neglected. We notice that some  studies have focused on selecting hard negative samples for IR tasks~\cite{DBLP:conf/sigir/ZhanM0G0M21,DBLP:conf/naacl/QuDLLRZDWW21,DBLP:conf/emnlp/LiTWFZY19}, but they did not use CL. In this paper, we propose two complementary and contrastive curricula to enhance the model's learning using both positive and negative context-document pairs. Our experiments show that regulating the learning pace of both positive and negative samples is very effective.



\begin{figure*}
    \centering
    \includegraphics[width=.95\linewidth]{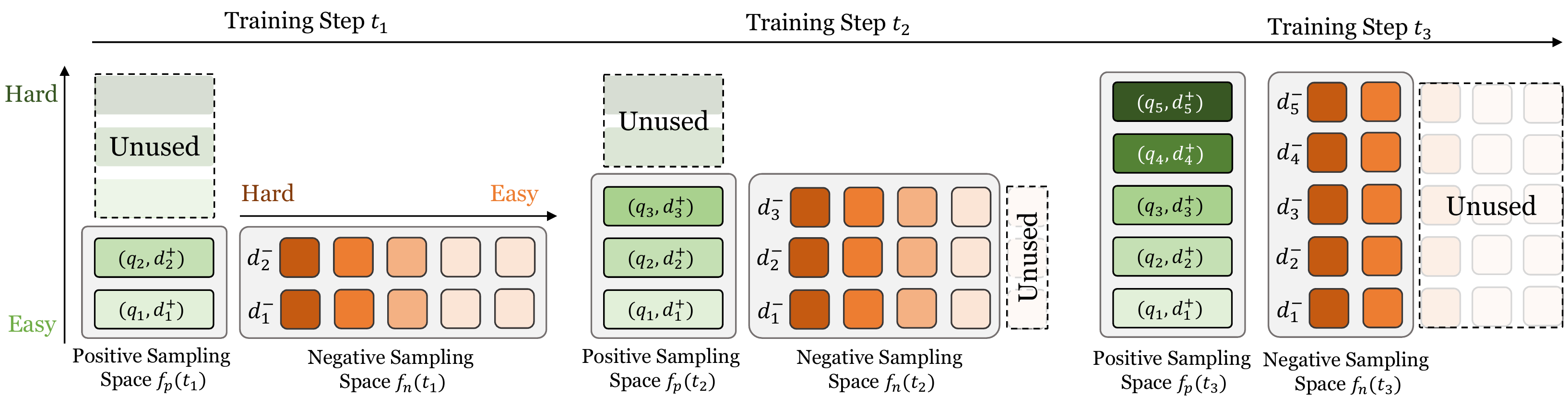}
    \caption{The training process of our framework. For the curriculum of \textcolor[rgb]{0,0.75,0}{positive pairs}, only easy samples are used at the beginning ($t_1$). Along the training process, the positive sampling space is gradually extended to the whole positive pairs ($t_3$). For the curriculum of \textcolor[rgb]{1,0.5,0}{negative pairs}, the sampling space is shrinking from all samples ($t_1$) to only hard samples ($t_3$).}
    \label{fig:framework}
    \vspace{-10px}
\end{figure*}

\section{Methodology}
The goal of context-aware document ranking is to rank a list of candidate documents using the search context. In this work, we propose a \textbf{D}ual \textbf{C}urriculum \textbf{L}earning (\ttt{DCL}) framework for this task. As shown in Figure~\ref{fig:method}, our framework consists of two complementary curricula for learning positive and negative context-document pairs. In each curriculum, the pairs are sorted according to their difficulty so that the model can learn them from easy to hard.

\subsection{Notations and Task Definition}
We first define some important concepts and notations before introducing our framework. A user's search behavior is represented as a sequence of $n$ interactions $H_n = [q_1,d_1^+\cdots,q_n,d_n^+]$, where each query $q_i$ is associated with 
a corresponding clicked document $d_i^+$. If there are several clicked documents, each of them is associated with the query to form a separate pair in the sequence. 
Each query $q_i$ is represented by the original text string submitted to the search engine, while each clicked document $d_i$ is represented by its text content. All queries are ordered according to the timestamps. For convenience, we further denote $C=[q_1,d_1^+,\cdots,q_n]$ as a \textbf{search context} when $q_n$ is submitted after a series of queries and interacted documents. When training a ranking model, for each positive clicked document, a set of $m$ unclicked documents are  selected as negative candidates.\footnote{Different selection strategies of negative candidates can be used in different datasets.} As a result, the candidate document set for the search context $C$ is represented as $D=\{d_n^+, d_1^-, \cdots\, d_{m}^-\}$ (the subscript $n$ in $d_n^+$ will be omitted).

With the above notations, the task of context-aware document ranking can be defined as: ranking the candidate document set $D$ based on the search context $C$ so as to rank the clicked document $d^+$ as high as possible. 


To make it clear, henceforth, we will call the pair $(C, d^+)$ a \textbf{positive} pair, whereas the pairs $(C, d_i^{-})_{i=1}^{m}$ are $m$ \textbf{negative} pairs.

Notice that in this paper, we leverage user logs to learn a ranking model because user clicks can serve as a good proxy of relevance feedback~\cite{DBLP:conf/sigir/JoachimsGPHG05,DBLP:journals/tois/JoachimsGPHRG07,CARS,HBA}. 
However, the same method can be used with human relevance judgments if available.

\subsection{Overview}
Our framework consists of two complementary curricula as follows:

(1) \textbf{Curriculum of Positive Pairs.}
As shown in the left side of Figure~\ref{fig:method}, our first curriculum is designed for positive pairs. The target is to teach the model how to understand users' intent by capturing matching clues between search context and clicked documents. To achieve this goal, we first sort all positive pairs in the training set according to their difficulty and then let the model learn from easy ones to hard ones. 
By this means, the model can gradually increase its ability to capture matching signals. 

(2) \textbf{Curriculum of Negative Pairs.}
We also design a curriculum for negative pairs (right side of  Figure~\ref{fig:method}) to enhance the model's ability to identify the mismatching information between search context and negative documents. Specifically, we progressively increase the difficulty of the negative candidate documents \textbf{associated with each selected positive pair} in the training set. In this way, the model is  encouraged to gradually distinguish more subtle mismatching signals between search context and negative documents.

The principles used in the two curricula have some differences. 
The learning from positive pairs aims at identifying the relevance signals. In the curriculum, the model can gradually capture signals from shallow and easy matching (such as similar terms) to deep and hard matching (such as semantic information). 
On the contrary, a group of negative documents is associated with a specific positive pair, so they provide supplementary mismatching signals. As the curriculum progresses, the negative document becomes more similar to the positive document, so the model has to capture more fine-grained clues to distinguish them. 
Overall, we train the model with the two complementary curricula simultaneously so that its capability of modeling users' intent can be gradually enhanced. The whole process is shown in Figure~\ref{fig:framework}.

It is worth noting that our \ttt{DCL} is a general training framework that works by organizing the learning order of the training samples. Therefore, it can be applied to various base models to improve their performance (this will be shown in our experiments).

\subsection{Dual Curriculum Learning Framework}
The implementation of our approach involves several key concepts, which we examine below.

\textit{How does curriculum learning work?} When training neural networks, a mini-batch of training samples is usually randomly (\ie, uniformly -- every sample is selected with the same likelihood) sampled from the training set and used for optimizing the model at a step. Curriculum learning, on the other hand, aims at adjusting the order of the samples so that they are learned according to a predefined pace rather than at random. In our framework, we design two curricula for learning positive pairs and negative pairs from easy to hard, respectively. Following the paradigm of curriculum learning~\cite{DBLP:conf/icml/HacohenW19,DBLP:journals/corr/abs-1912-08555}, each curriculum is defined by two functions:

$\bullet$ A \textit{difficulty} function determines the difficulty of samples so that the samples can be sorted according to their difficulty.

$\bullet$ A \textit{pacing} function controls the learning pace. Essentially, it adjusts the sampling space to control the difficulty of the training samples at each step. 

The curricula and their combination are described below.


\subsubsection{Curriculum of Positive Pairs}\label{sec:cpp}

The green part of Figure~\ref{fig:framework} illustrates the curriculum of positive pairs. Positive pairs are sorted from easy to hard according to a difficulty function. Thereafter, we gradually enlarge the sampling space so that more difficult positive samples will be included.
The key is to define an appropriate difficulty function.

\textbf{Difficulty Function.} Different heuristics can be used to determine the difficulty of documents for a query \cite{DBLP:conf/sigir/MacAvaneyN0TGF20a,DBLP:conf/sigir/CarmelY10}. 
We consider two factors for measuring the difficulty of each positive training pair $(C, d^+)$: (1) The first factor is the ranking score $M(\cdot,\cdot)$ between $C$ and $d^+$ (see details in Section~\ref{sec:rm}): A higher $M(C,d^+)$ indicates that 
it is easier to select $d^+$ based on $C$.
(2) We also consider the position of the clicked document in the ranked list: A higher position indicates that it is easier to  select it out of all documents.
Formally, the difficulty for $(C, d^+)$ is computed as follows:
\begin{align}
    d_p(C, d^+) = \underbrace{\operatorname{rank}_C(d^+)}_{\in [1, \lvert\mathcal{D}\rvert]} + \underbrace{\left(1 - \frac{M(C,d^+)}{\max_{(C_i, d_i^+)\in \mathcal{D}}M(C_i,d_i^+)}\right)}_{\in (0, 1]},\label{eq:dp}
\end{align}
where $\mathcal{D}$ is the training set. The first term is the ranking position of $d^+$ under the search context $C$; the second term is the normalized ranking score of $(C,d^+)$. In this function, the difficulty is dominated by the position of the clicked document (the first term), while the normalized ranking score (the second term) makes an effect only when different $(C,d^+)$ pairs are ranked at the same position. This particular definition of difficulty leads to good experimental performance among different alternatives we tested. 

\begin{figure}
    \centering
    \includegraphics[width=.73\linewidth]{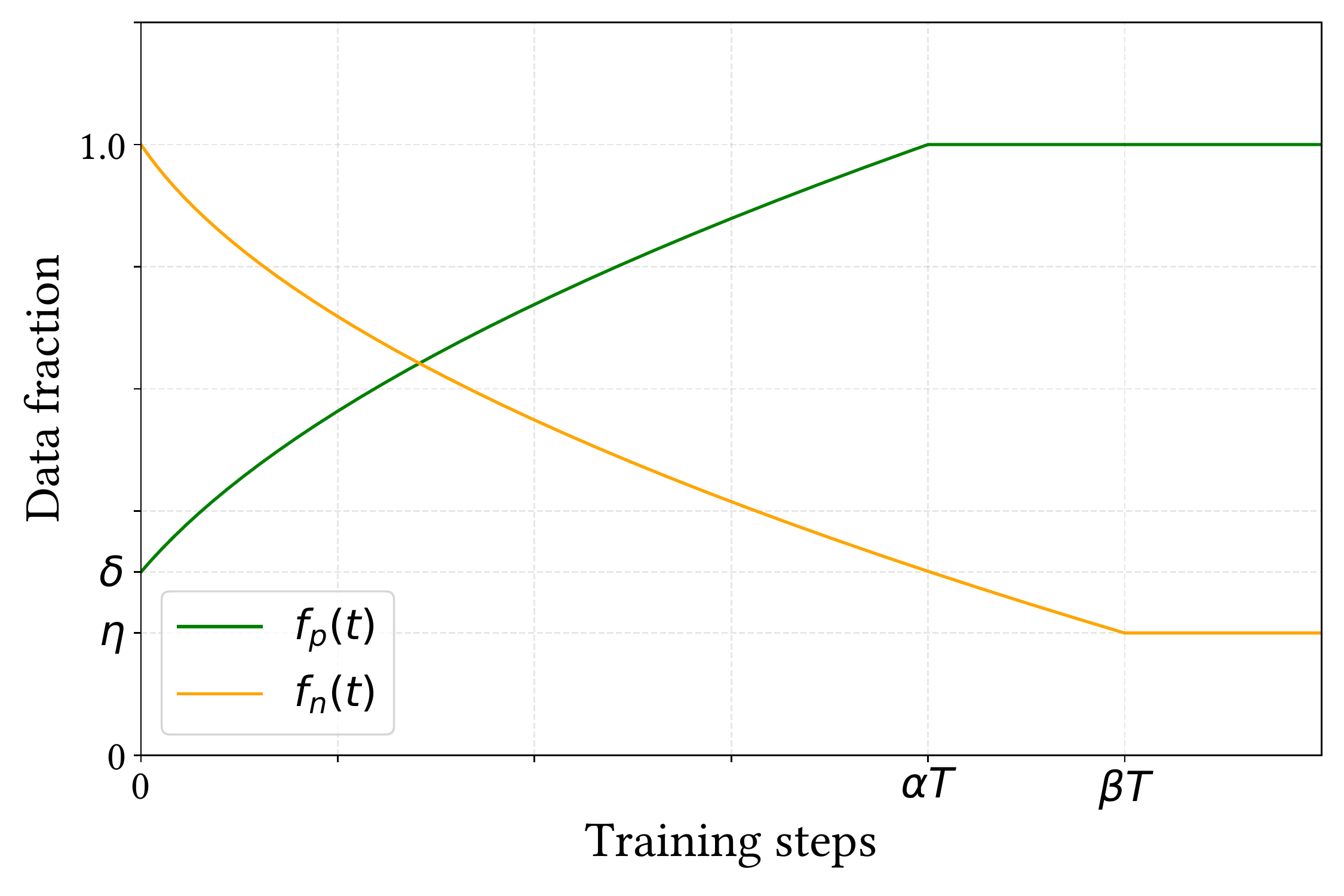}    
     \caption{Pacing functions used in both curricula.}
    \label{fig:f}
    \vspace{-10px}
\end{figure}

\textbf{Pacing Function.}
The pacing function determines how the training process transitions from easy to hard pairs. Following previous studies on curriculum learning~\cite{DBLP:journals/corr/abs-1912-08555,DBLP:conf/icml/HacohenW19}, we define the pacing function $f_p(t)$ with respect to the training step $t$. The value of $f_p(t)$ is a proportion, and only the first $f_p(t)\times\lvert\mathcal{D}\rvert$ positive pairs can be used at the training step $t$. $f_p(t)$ is defined as follows:
\begin{align}
    f_p(t) = \min\left(1.0, \left(t\cdot\frac{1-\delta^k}{\alpha T}+\delta^k\right)^{\frac{1}{k}}\right),\label{eq:fp}
\end{align}
where $T$ is the total number of training steps, $\alpha,\delta\in(0,1)$ and $k\in[1,+\infty)$ are hyperparameters. As shown by the green line in Figure~\ref{fig:f}, this function has the following properties: (1) the initial value $f_p(0)$ is $\delta$, so that the model can only use some easy pairs in the first training step; (2) it increases monotonically so that harder pairs are added to the training set gradually; (3) when it reaches $\alpha T$ steps ($f_p(\alpha T)=1$), all pairs in the corpus can be used for training. 


\subsubsection{Curriculum of Negative Pairs}\label{sec:cnp} 
Negative samples are also very important for learning a ranking model~\cite{DBLP:conf/sigir/ZhanM0G0M21}. By comparing positive and negative pairs, the model can learn what matching signals are vital in a contrastive manner. 

The orange part of Figure~\ref{fig:framework} shows the curriculum of negative pairs. Similar to positive pairs, the negative pairs are also arranged according to their difficulties (more details later). By gradually constraining the sampling space, the model will focus on more difficult samples in later steps.

\textbf{Difficulty Function.} Similar to the positive curriculum, we use the relevance between the search context and the negative candidate as the difficulty of the negative pair. A negative candidate with a high ranking score to the search context is deemed to be hard to distinguish. For a negative pair $(C,d^-_i)$, its difficulty is defined as:
\begin{align}
    d_n(C,d^-_i) = M(C,d^-_i),\label{eq:dn}
\end{align}
where $M(\cdot,\cdot)$ is a scoring model similar to that used in the curriculum of positive pairs. Different from the difficulty function $f_p(t)$ in Equation~\ref{eq:fp}, it is unnecessary to introduce the ranking position and normalization operation, because all negative candidates are associated with the same query; their ranking positions are determined by the ranking scores. 


It is worth noting that we follow previous studies~\cite{CARS} to select negative candidates: the unclicked documents ranked around the clicked document (within a window) are considered as negative samples. By this means, we can avoid using too trivial or too hard negative documents. The obtained list of negative pairs is denoted as $\mathcal{L}$, in which all pairs are sorted according to their difficulties descendingly.


\textbf{Pacing Function.} The pacing function $f_n(t)$ for negative pairs is designed in a similar way to that for positive pairs. 
$f_n(t)$ adjusts the sampling space from which the negative sequences are sampled. It decreases with $t$. At time step $t$, we sample from the first $f_n(t)\times \lvert\mathcal{L}\rvert$ negative pairs for training. 
$f_n(t)$ is defined as:
\begin{align}
    f_n(t) = \max\left(\eta, 1 + \eta - \left(t\cdot\frac{1-\eta^k}{\beta T}+\eta^k\right)^{\frac{1}{k}}\right),\label{eq:fn}
\end{align}
where $\beta,\eta\in(0,1)$ are hyperparameters. As shown by the organce line in Figure~\ref{fig:f}, this function is similar to $f_p(t)$, but makes opposite effect (\ie, decreasing rather than increasing) along with $t$. 

\textbf{Remark.} As shown in Figure~\ref{fig:framework} and Figure~\ref{fig:f}, we design the curriculum of negative pairs in a manner opposite to that of positive pairs, namely we gradually focus on only hard negative pairs. This is because: (1) All positive pairs are collected from human click data, which are very valuable for learning the real user intent. So, we choose to enlarge the positive sampling space and use all positive pairs in the end. (2) The negative pairs are sampled from the repository for facilitating the learning of the associating positive pairs. When the training progresses, too easy samples cannot provide enough ``contrast effect'', thus we discard them and only focus on hard samples. This strategy can lead to a more robust optimization in retrieval performance~\cite{DBLP:conf/sigir/ZhanM0G0M21}.

\begin{algorithm}[!t]
	\caption{Training in \ttt{DCL}}
	\small
	\begin{algorithmic}[1]\label{alg:clef}
	    \STATE \textbf{Input:} the dataset $\mathcal{D}$; the ranking model; difficulty functions $d_p(\cdot,\cdot)$ and $d_n(\cdot,\cdot)$; pacing functions $f_p(\cdot)$ and $f_n(\cdot)$; the number of negative pairs $m$. 
	    \STATE Score all positive pairs by $d_p(C,d^+)$ and sort them ascendingly;
	    \FOR{each training step $t$}
	    \STATE Collect the first $f_p(t)\cdot\lvert\mathcal{D}\rvert$ positive pairs as a subset $\mathcal{P}$;
        \STATE Uniformly sample a batch of positive pairs $\mathcal{B}_t$ from $\mathcal{P}$;
        \FOR{$[C_i,d^+_i]$ in $\mathcal{B}_t$}
        \STATE Score all negative pairs by $d_n(C_i,d^-_j)$, get a list of negative pairs $\mathcal{L}$ by the heuristic rule, and sort them descendingly;
        \STATE Collect the first $f_n(t)\cdot\lvert\mathcal{L}\rvert$ negative pairs as a subset $\mathcal{N}$;
        \STATE Uniformly sample $m$ pairs $D_i^-$ from $\mathcal{N}$;
        \ENDFOR
        \STATE Optimize the ranking model on the data $\{C_i,d_i^+,D_i^-\}_{i=1}^{\lvert \mathcal{B}_t\rvert}$;
        \ENDFOR
        \STATE \textbf{Output:} Trained ranking model.
	\end{algorithmic}
\end{algorithm}

\subsubsection{Combination of Two Curricula} 
\ttt{DCL} trains ranking models with the two curricula simultaneously. Specifically, for a training step $t$, we build a batch of training data as follows:
First, we select a batch of positive pairs $(C, d^+)$ according to the pacing function $f_p(t)$. Then, for each search context $C$ in the positive pairs, we sample $m$ negative candidate documents based on the pacing function $f_n(t)$. The process is summarized in Algorithm~\ref{alg:clef}.

The whole training process of our framework naturally simulates two learning paradigms of human beings. On the one hand, the learning material is organized from easy to hard, which has been demonstrated to be effective for both animal training and human learning. In our case, such a training process is beneficial for model optimization. On the other hand, in cognitive science, learning through comparison is also an effective way to understand new concepts~\cite{DBLP:conf/cogsci/HigginsR11}. In our framework, the model can learn by contrasting positive and negative candidate documents. 

\subsubsection{Selection of Scoring Model}\label{sec:rm}
In Equation~(\ref{eq:dp}) and~(\ref{eq:dn}), \ttt{DCL} applies a scoring model $M(\cdot,\cdot)$ to measure the ranking score between the search context and the candidate document. In our experiments, we use  two different methods for $M(\cdot,\cdot)$: (1) BM25~\cite{BM25} and (2) a BERT-based score commonly used in dense retrieval~\cite{DBLP:conf/emnlp/KarpukhinOMLWEC20,DBLP:conf/sigir/KhattabZ20}. Since we need to compute the ranking score between a search context and all documents in the training set, we apply dot-product based on BERT representations for fast computation: 
\begin{align}
    M(C, d) = {\operatorname{BERT}}(C)_{\rm [CLS]} \cdot {\operatorname{BERT}}(d)_{\rm [CLS]}.
\end{align}
To achieve better performance, we fine-tune BERT encoders on positive sequences with in-batch negatives~\cite{DBLP:conf/emnlp/KarpukhinOMLWEC20}. Then, we can obtain the representations of all the search contexts and documents. Afterwards, we use FAISS~\cite{DBLP:journals/tbd/JohnsonDJ21} to compute $M(C, d)$ efficiently. 

\begin{table}[t!]
    \centering
    \small
    \caption{The statistics of the datasets. The number in paretheses is the average number of relevant documents.}
    \begin{tabular}{lrrrrrr}
    \toprule
        \textbf{AOL} & \textbf{Training} & \textbf{Validation} & \textbf{Test} \\
    \midrule
        \# Sessions & 219,748 & 34,090 & 29,369 \\
        \# Queries  & 566,967 & 88,021 & 76,159 \\
        Avg. \# Query per Session & 2.58 & 2.58 & 2.59 \\
        Avg. \# Document per Query & 5 & 5 & 50 \\
        Avg. Query Len & 2.86 & 2.85 & 2.9 \\
        Avg. Document Len & 7.27 & 7.29 & 7.08 \\   
        Avg. \# Clicks per Query & 1.08 & 1.08 & 1.11 \\
    \midrule
        \textbf{Tiangong-ST} & \textbf{Training} & \textbf{Validation} & \textbf{Test} \\
    \midrule
        \# Sessions & 143,155 & 2,000 & 2,000 \\
        \# Queries  & 344,806 & 5,026 &  6,420 \\
        Avg. \# Query per Session & 2.41 & 2.51 & 3.21 \\
        Avg. \# Document per Query & 10 & 10 & 10 \\
        Avg. Query Len & 2.89 & 1.83 & 3.46 \\
        Avg. Document Len & 8.25 & 6.99 & 9.18 \\
        Avg. \# Clicks per Query & 0.94 & 0.53 & (3.65) \\
    \bottomrule
    \end{tabular}
    \vspace{-10px}
    \label{tab:sta}
\end{table}

\section{Experiments}

\begin{table*}[t!]
    \centering
    \small
    \caption{Experimental results on two datasets. All results  using our framework (\ttt{X+DCL}) outperforms the original results (\ttt{X}) significantly at $p$-value $<$ 0.01 with Bonferroni correction in paired t-test.}
    \setlength{\tabcolsep}{1.4mm}{
    \begin{tabular}{lcccccccccccc}
    \toprule
        & \multicolumn{6}{c}{AOL} & \multicolumn{6}{c}{Tiangong-ST} \\
        \cmidrule(lr){2-7}\cmidrule(lr){8-13}
        Model & MAP & MRR & NDCG@1 & NDCG@3 & NDCG@5 & NDCG@10 & MAP & MRR & NDCG@1 & NDCG@3 & NDCG@5 & NDCG@10 \\
    \midrule
        \ttt{ARC-I} & 0.3361 & 0.3475 & 0.1988 & 0.3108 & 0.3489 & 0.3953 & 0.8580 & 0.9159 & 0.7088 & 0.7087 & 0.7317 & 0.8691 \\
        \ttt{ARC-II} & 0.3834 & 0.3951 & 0.2428 & 0.3564 & 0.4026 & 0.4486 & 0.8611 & 0.9227 & 0.7131 & 0.7237 & 0.7379 & 0.8732 \\
        \ttt{KNRM} & 0.4038 & 0.4133 & 0.2397 & 0.3868 & 0.4322 & 0.4761 & 0.8709 & 0.9261 & 0.7473 & 0.7505 & 0.7624 & 0.8891 \\
        \ttt{Duet} & 0.4008 & 0.4111 & 0.2492 & 0.3822 & 0.4246 & 0.4675 & 0.8663 & 0.9273 & 0.7577 & 0.7354 & 0.7548 & 0.8829 \\
        \ttt{M-NSRF} & 0.4217 & 0.4326 & 0.2737 & 0.4025 & 0.4458 & 0.4886 & 0.8517 & 0.9084 & 0.7124 & 0.7308 & 0.7489 & 0.8795 \\
        \ttt{M-Match} & 0.4459 & 0.4572 & 0.3020 & 0.4301 & 0.4697 & 0.5103 & 0.8529 & 0.9211 & 0.7311 & 0.7233 & 0.7427 & 0.8801 \\
        \ttt{CARS} & 0.4297 & 0.4408 & 0.2816 & 0.4117 & 0.4542 & 0.4971 & 0.8556 & 0.9268 & 0.7385 & 0.7386 & 0.7512 & 0.8837 \\
        \midrule
        \ttt{HBA} & 0.5281 & 0.5384 & 0.3773 & 0.5241 & 0.5624 & 0.5951 & 0.8615 & 0.9316 & 0.7612 & 0.7518 & 0.7639 & 0.8896 \\
        \ttt{HBA+DCL} & 0.5599 & 0.5693  & 0.4074 & 0.5626 & 0.5961 & 0.6242 & 0.8986 & {0.9538} & {0.8069} & {0.7985} & {0.8130} & {0.9122} \\
        {Improv.} & +6.02\% & +5.74\% & +7.98\% & +7.35\% & +5.99\% & +4.89\% & +4.31\% & +2.38\% & +6.00\% & +6.21\% & +6.43\% & +2.54\% \\
        \midrule
        \ttt{RICR} & 0.5338 & 0.5450 & 0.3894 & 0.5267 & 0.5648 & 0.5971 & 0.8147 & 0.8937 & 0.7670 & 0.7636 & 0.7740 & 0.8934 \\
        \ttt{RICR+DCL} & 0.5630 & 0.5742 & 0.4219 & 0.5589 & 0.5943 & 0.6257 & 0.8963 & 0.9498 & 0.7995 & 0.7925 & 0.8078 & 0.9089 \\
        {Improv.} & +5.47\% & +5.36\% & +8.35\% & +6.11\% & +5.22\% & +4.79\% & +10.02\% & +6.28\% & +4.24\% & +3.78\% & +4.37\% & +1.73\% \\
        \midrule
        \ttt{COCA} & 0.5500 & 0.5601 & 0.4024 & 0.5478 & 0.5849 & 0.6160 & 0.8623 & 0.9382 & 0.7769 & 0.7576 & 0.7703 & 0.8932 \\
        \ttt{COCA+DCL} & {0.5794} & {0.5888} & {0.4281} & {0.5841} & {0.6167} & {0.6432} & {0.8990} & {0.9501} & {0.7936} & {0.7922} & {0.8077} & {0.9088} \\
        {Improv.} & +5.35\% & +5.12\% & +6.38\% & +6.63\% & +5.44\% & +4.42\% & +3.21\% & +1.27\% & +2.15\% & +4.57\% & +4.86\% & +1.75\% \\
    \bottomrule
    \end{tabular}
    }
    \label{tab:result}
    \vspace{-10px}
\end{table*}

\subsection{Datasets and Evaluation Metrics}
Following previous work~\cite{HBA,CARS,RICR,COCA}, we conduct experiments on two public datasets: AOL search log data~\cite{AOL} and Tiangong-ST query log data~\cite{Tiangong}.\footnote{We understand that the AOL dataset should normally not be used in experiments. We still use it because there are not many datasets available that fit our experiments well.} Another possibility is the MS MARCO Conversational Search dataset~\cite{DBLP:conf/nips/NguyenRSGTMD16}, but the sessions in it are artificially constructed rather than derived from real search logs and the corresponding clicked documents are unavailable. Therefore, we do not use the MS MARCO dataset in experiments.

We use the \textbf{AOL} dataset constructed by~\citet{CARS}. The dataset contains a large number of sessions, each of which consists of several queries. Each query in the training and validation sets has five candidate documents. The test set uses 50 documents retrieved by BM25~\cite{BM25} as candidates for each query. All queries have at least one satisfied click in this dataset.

\textbf{Tiangong-ST} dataset is collected from a Chinese commercial search engine. It contains web search session data extracted from an 18-day search log. Each query in the dataset has ten candidate documents. In the training and validation sets, we use the clicked documents as the satisfied clicks. For queries with no satisfied clicks, we use a special token ``[Empty]'' for padding. In the test set, the candidate documents for the last query of each session have been manually annotated with relevance scores from from 0 to 4, which are used for evaluation. More details can be found in~\cite{Tiangong}. 





The statistics of both datasets are shown in Table~\ref{tab:sta}. In Tiangong-ST test set, the relevance scores are provided instead of click labels, so we report the average number of documents with positive relevance scores ($\geq$ 1). Following previous studies~\cite{DBLP:conf/cikm/HuangHGDAH13,DBLP:conf/ijcai/HuangZSWL18,CARS,COCA}, to reduce memory requirements and speed up training, we only use the document title as its content.

\textbf{Evaluation Metrics.} Similar to previous studies~\cite{CARS,HBA,COCA}, we use Mean Average Precision (MAP), Mean Reciprocal Rank (MRR), and Normalized Discounted Cumulative Gain at position $k$ (NDCG@$k$, $k=\{1,3,5,10\}$) as evaluation metrics. In AOL, as clicked documents are used instead of manually judged documents, these measures reflect the capability of a model to rank clicked documents high. 
All evaluation results are obtained using the TREC's standard evaluation tool (trec\_eval)~\cite{DBLP:conf/sigir/GyselR18}. 

\subsection{Baseline}
We compare our method with several baseline methods, including those for (1) ad-hoc ranking and (2) context-aware ranking.

(1) \textbf{Ad-hoc ranking methods} only use the current query without context information (historical queries and documents) for document ranking.

$\blacktriangleright$~\ttt{ARC-I}~\cite{ARC-I} is a representation-based approach. Query and  document representations are generated by CNNs. The ranking score is determined by a multi-layer perceptron (MLP).
$\blacktriangleright$~\ttt{ARC-II}~\cite{ARC-I} is an interaction-based method. A matching map is constructed from the query and document, from which CNNs extract matching features. The score is also computed by an MLP.
$\blacktriangleright$~\ttt{KNRM}~\cite{KNRM} constructs a matching matrix by performing fine-grained interaction between the query and documents. The ranking features and scores are computed via kernel pooling.
$\blacktriangleright$~\ttt{Duet}~\cite{DUET} uses both interaction- and representation-based features of the query and document extracted by CNNs and MLPs to compute ranking scores.

(2) \textbf{Context-aware ranking methods}  utilize both context information and the current query to rank candidate documents.

$\blacktriangleright$~\ttt{M-NSRF}~\cite{MNSRF} is a multi-task model that jointly predicts the next query and ranks corresponding documents. An RNN encodes a session's historical queries. The ranking score is calculated based on the representation of the query, the history, and the document.
$\blacktriangleright$~\ttt{M-Match-Tensor}~\cite{MNSRF} (henceforth denoted as \ttt{M-Match}) is similar to \ttt{M-NSRF}, but learns a contextual representation for each word in the queries and documents. The ranking score is calculated by word-level representation.
$\blacktriangleright$~\ttt{CARS}~\cite{CARS} also learns query suggestion and document ranking simultaneously. An attention mechanism is applied to compute representations for each query and document. The final ranking score is computed using the representation of historical queries, clicked documents, current query, and candidate documents.\footnote{We will notice some slight discrepancies between our results and those of the original paper of CARS. This is due to different tie-breaking strategies in evaluation. Following~\cite{HBA,COCA}, we use trec\_eval while the authors of CARS use their own implementation.}
$\blacktriangleright$~\ttt{HBA-Transformer}~\cite{HBA} (henceforth denoted as \ttt{HBA}) concatenates historical queries, clicked documents, and unclicked documents into a long sequence and applies BERT~\cite{BERT} to encode them into representations. A higher-level transformer structure with behavior embedding and relative position embedding enhances the representation. Finally, the ranking score is computed based on the representation of the ``[CLS]'' token.
$\blacktriangleright$~\ttt{RICR}~\cite{RICR} is a unified context-aware document ranking model which takes full advantage of both representation and interaction. The session history is encoded into a latent representation and used to enhance the current query and the candidate document. Several matching components are applied to capture the interaction between the enhanced query and candidate documents. This model is based on RNNs and attention mechanism.
$\blacktriangleright$~\ttt{COCA}~\cite{COCA} uses contrastive learning to improve a BERT's representation of user behavior sequences. By distinguishing similar user behavior sequences with dissimilar ones, the encoder can generate more robust representation. Then, the encoder is further used in context-aware document ranking. This is the current state-of-the-art document ranking method based on user behavior sequence.

Due to the limited space, the \textbf{implementation details} are provided in our code repository.\footnote{\url{https://github.com/DaoD/DCL}}

\subsection{Experimental Results}
Experimental results are shown in Table~\ref{tab:result}. We choose three recently proposed methods (\ie, \ttt{HBA}, \ttt{RICR}, and \ttt{COCA}) as the base model and train them with our proposed \ttt{DCL} framework. The corresponding results are reported as ``\ttt{X+DCL}''. To avoid the influence of randomness, we set three different random seeds and report the average performance. The standard deviation is less than 1e-3 for all results, which is omitted in the table. 
In general, our \ttt{DCL} significantly improves the performance of three base models in terms of all evaluation metrics on both datasets. This result clearly demonstrates \ttt{DCL}'s superiority. We also have the following observations.

(1) The context-aware document ranking models generally perform better than ad-hoc ranking methods. For instance, on the AOL dataset, the weak contextualized model \ttt{M-NSRF} can still outperform the strong ad-hoc ranking model \ttt{KNRM}. This indicates that modeling user historical behavior is beneficial for understanding the user intent and determining the desired documents. For the three strong context-aware models, \ttt{DCL} can further improve their performance greatly, showing its effective utilization of training samples via curriculum learning.

(2) Compared with RNN-based methods (such as \ttt{RICR} and \ttt{CARS}), BERT-based methods (\ttt{COCA} and \ttt{HBA}) perform better. It is noticeable that \ttt{DCL} can bring improvements for both kinds of methods. Specifically, it improves the results by more than 4.7\% and 1.2\% in terms of all metrics on AOL and Tiangong-ST, respectively.  This result demonstrates the wide applicability of our method to different base models and reflects the evident advantage of learning samples from easy to hard in context-aware document ranking.

(3) \ttt{COCA} is the state-of-the-art approach to context-aware document ranking. It involves a contrastive learning pre-training stage to help the BERT encoder learn more robust representations for user behavior sequences. In comparison, our \ttt{DCL} is a general training framework without a specific pre-training step, making it more efficient in practice.\footnote{Compared to the pre-training step in \ttt{COCA}, \ttt{DCL} takes only around 1/3 time for training a BERT scorer, and this cost can be further reduced if BM25 scorer is applied.} 
In addition,  \ttt{COCA} trained with \ttt{DCL} achieves a new state-of-the-art performance in context-aware document ranking task, showing the usefulness to combine  both curriculum learning and contrastive learning.

\begin{table}[t!]
    \centering
    \small
    \caption{Ablation study of \ttt{COCA+DCL}. Mark ``$\checkmark$'' and ``$\times$'' indicate whether a curriculum is used or not. ``Easy'' or ``hard'' means only easy/hard samples are used for training.}
    \setlength{\tabcolsep}{1.3mm}{
    \begin{tabular}{ccccccc}
        \toprule
        Pos. & Neg. & MAP & MRR & NDCG@1 & NDCG@3 & NDCG@10 \\
        \midrule
        $\times$ & $\times$ & 0.5500 & 0.5601 & 0.4024 & 0.5478 & 0.6160 \\
        $\checkmark$ & $\times$ & 0.5750 & 0.5843 & 0.4222 & 0.5811 & 0.6391 \\
        $\times$ & $\checkmark$ & 0.5740 & 0.5843 & 0.4231 & 0.5791 & 0.6381 \\
        $\checkmark$ & $\checkmark$ & \textbf{0.5794} & \textbf{0.5888} & \textbf{0.4281} & \textbf{0.5841} & \textbf{0.6432} \\
        \midrule
        $\times$ & Easy & 0.5240 & 0.5350 & 0.3677 & 0.5200 & 0.5939 \\
        $\times$ & Hard & 0.5698 & 0.5792 & 0.4173 & 0.5741 & 0.6339 \\
        \bottomrule
    \end{tabular}
    }
    \label{tab:ab}
    \vspace{-10px}
\end{table}

\subsection{Discussion}
We further discuss several aspects of our proposed \ttt{DCL}. These analyses are based on the results on the AOL dataset, while we have similar findings on the Tiangong-ST dataset.

\subsubsection{Impact of Both Curricula}
As we propose two curricula for learning on positive and negative pairs, to validate their effectiveness, we conduct an ablation study by disabling each of them from \ttt{COCA+DCL} (\ie, the sampling is done among all samples). The results are shown in Table~\ref{tab:ab}. We can observe:

First, we can see that both curricula are useful. Applying any of them leads to performance improvement. When no curriculum learning is used, we observe large drops in performance. This directly validates our assumption in this paper that learning from easy to hard samples can guide the model in a good learning direction. 
Second, the curriculum of positive pairs brings slightly higher improvement than that of negative pairs. This suggests that the ability to capture positive matching signals is more critical than being able to discard negative signals. A possible explanation is that positive signals are more focused while the negative ones are diffuse.

Furthermore, to investigate the influence of samples' difficulty changes during the training, we replace the curriculum of negative pairs by only using easy or hard pairs, and the curriculum of positive pairs is  disabled to avoid additional influence. The experimental results are shown in the bottom of Table~\ref{tab:ab}.

As can be seen, training with only hard negatives is even better than using all negatives (first row in the table). This finding is consistent with existing studies on using hard negatives to facilitate the optimization of dense retrievers~\cite{ance,DBLP:conf/naacl/QuDLLRZDWW21,DBLP:conf/sigir/ZhanM0G0M21}. However, only using easy samples for training makes the performance drop sharply. This is because the easy negatives cannot provide sufficient ``contrastive signals'' for learning the matching between search context and candidate documents. This is also why we design our curriculum of negative pairs as gradually shrinking to only hard negatives (details are presented in Section~\ref{sec:cnp}). Finally, dynamically adjusting the learning difficulty through curriculum is beneficial for model training (\eg, MAP is improved from 0.5698 to 0.5740). This demonstrates again the effectiveness of applying currciulum learning.

\begin{table}[t!]
    \centering
    \small
    \caption{Performance with different difficulty scorers.}
    \setlength{\tabcolsep}{1.6mm}{
    \begin{tabular}{llccccc}
        \toprule
        Pos. & Neg. & MAP & MRR & NDCG@1 & NDCG@3 & NDCG@10 \\
        \midrule
        \ttt{None} & \ttt{None} & 0.5500 & 0.5601 & 0.4024 & 0.5478 & 0.6160 \\
        \midrule
        \ttt{BM25} & \ttt{BM25} & 0.5661 & 0.5763 & 0.4159 & 0.5697 & 0.6293 \\
        \ttt{BM25} & \ttt{BERT} & 0.5600 & 0.5701 & 0.4081 & 0.5632 & 0.6244 \\
        \ttt{BERT} & \ttt{BM25} & \textbf{0.5794} & \textbf{0.5888} & \textbf{0.4281} & \textbf{0.5841} & \textbf{0.6432} \\
        \ttt{BERT} & \ttt{BERT} & 0.5652 & 0.5755 & 0.4152 & 0.5684 & 0.6290 \\
        \bottomrule
    \end{tabular}
    }
    \vspace{-10px}
    \label{tab:ranker}
\end{table}

\begin{figure}[t!]
    \centering
    \begin{subfigure}[b]{0.495\linewidth}
        \centering
        \includegraphics[width=\linewidth]{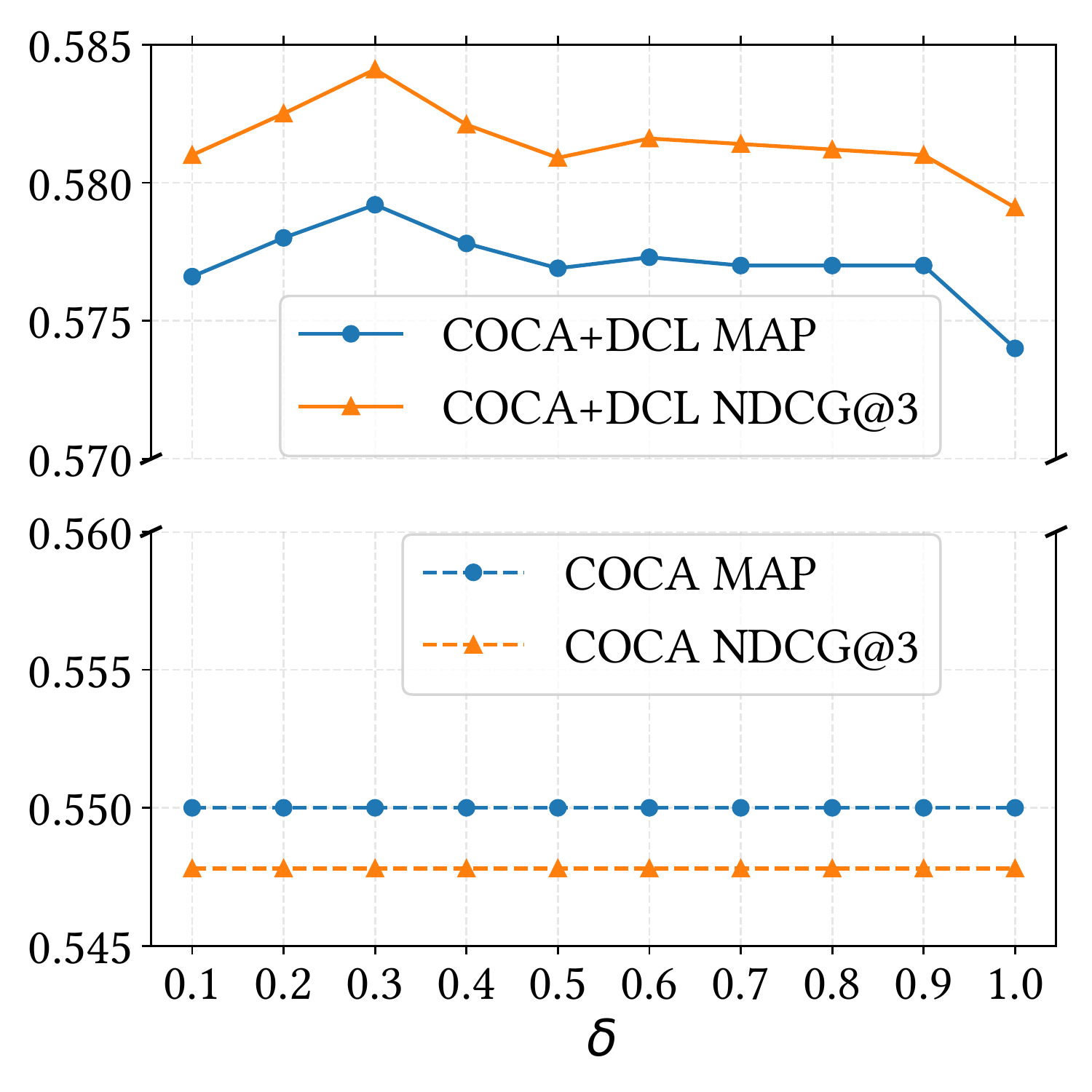}
    \end{subfigure}
    \begin{subfigure}[b]{0.495\linewidth}
        \centering
        \includegraphics[width=\linewidth]{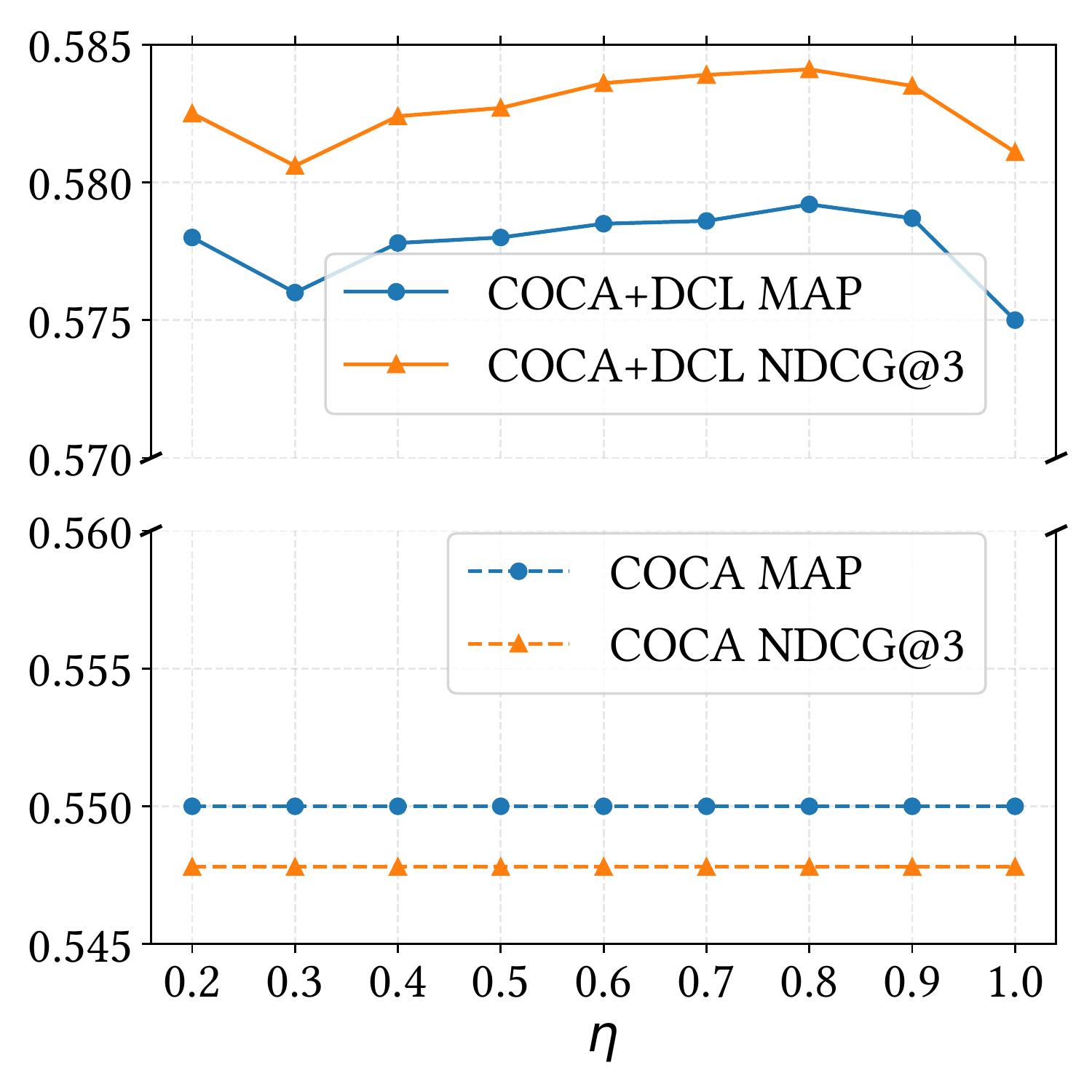}
    \end{subfigure}
    \caption{Performance with different hyperparameters.}
    \label{fig:hyper}
    \vspace{-10px}
\end{figure}

\begin{figure*}[t!]
    \centering
    \begin{subfigure}[b]{0.495\linewidth}
        \centering
        \includegraphics[width=\linewidth]{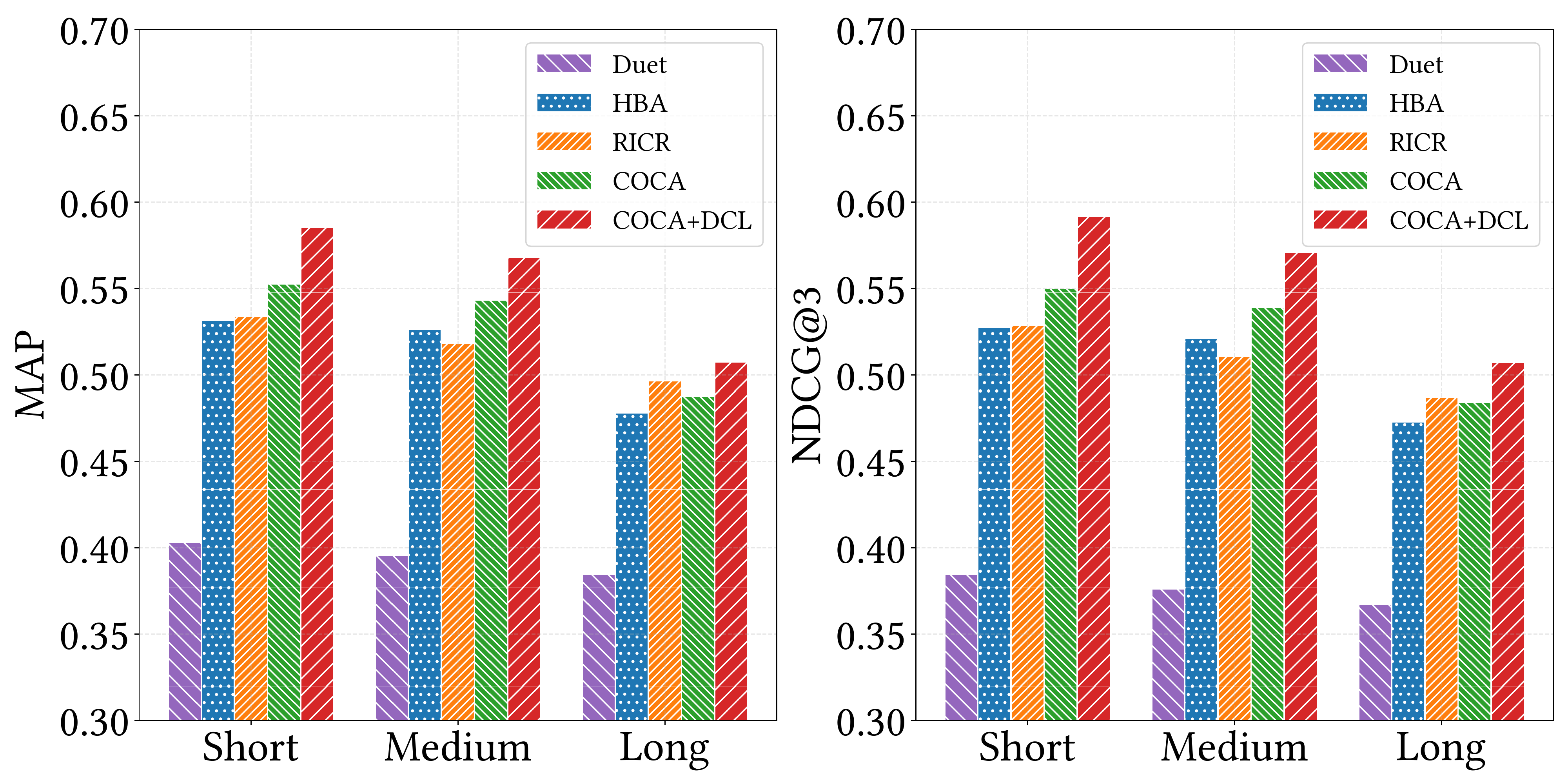}
    \end{subfigure}
    \begin{subfigure}[b]{0.495\linewidth}
        \centering
        \includegraphics[width=\linewidth]{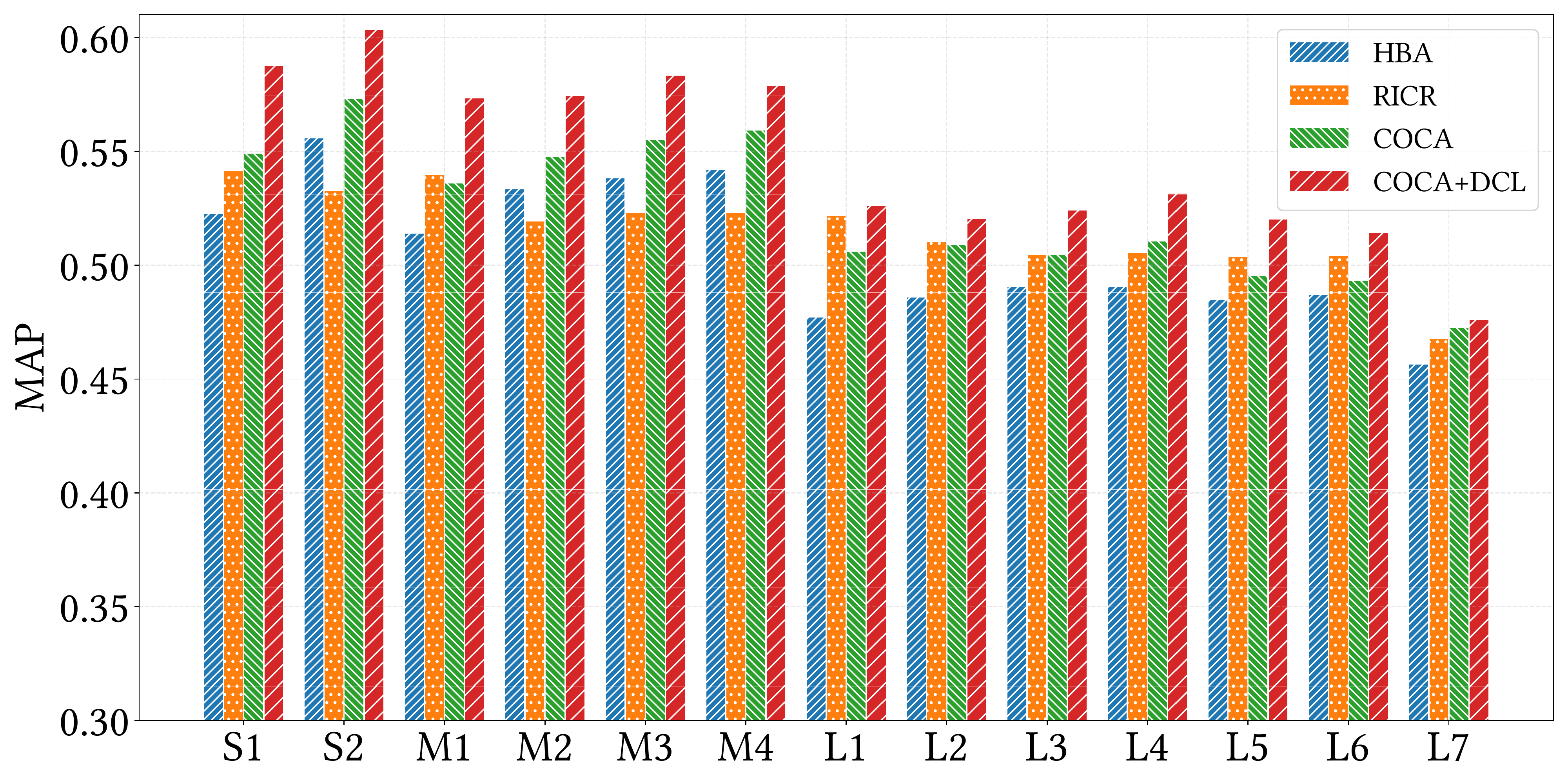}
    \end{subfigure}
    \caption{(Left) Performance on different lengths of sessions on AOL. (Right) Performance at different query positions in short (S1-S2), medium (M1-M4), and long sessions (L1-L7). The number after ``S'', ``M'', or ``L'' indicates the query index in the session.}%
    \vspace{-10px}
    \label{fig:len}
\end{figure*}

\subsubsection{Influence of Scoring Models}\label{sec:trm}
We proposed two scoring models for $M(\cdot,\cdot)$ -- BM25 and BERT. We investigate their impact, and Table~\ref{tab:ranker} shows the results of \ttt{COCA+DCL}. 

We can observe the following: (1) Despite the differences in performance, \ttt{DCL} combined with each of the scoring methods can consistently bring improvements over training without curriculum. This shows that both scoring functions can help determine the difficulty of pairs. 
(2) BERT scores work better on the positive pair curriculum, while BM25 scores are better on the negative pair curriculum.
The potential reasons could be: (a) The negative candidates provided in the test set are also selected by BM25. As a result, the distribution of negative pairs selected by BM25 on the training set is closer to that on the test set, allowing the model to perform better. (b) Negative pairs are used for learning mismatching signals. Compared with BERT, BM25 can identify negative candidates containing similar terms. As term-level matching signals are critical in IR, such negative candidates can provide more useful information on term-level dissimilarity.

\subsubsection{Influence of Hyperparameters}\label{sec:hyp}
In \ttt{DCL}, $\delta$ (in Equation~\ref{eq:fp}) and $\eta$ (in Equation~\ref{eq:fn}) are two hyperparameters that control the degree of difficulty in the initial set of positive pairs and in the set of final negative pairs. They are  determined according to the validation set. We show their impact in Figure~\ref{fig:hyper}. 

As we can see, when $\delta$ is small ($<0.4$, \ie, using very easy positive pairs at beginning), the performance is high. When $\delta$ becomes large (\ie, including more difficult positive pairs at beginning), the performance drops. This confirms our hypothesis that the optimization process can be confused with more difficult pairs at beginning. 
When $\delta=1.0$, the curriculum of positive pairs is disabled, so all positive pairs are learned in a random order. We can see that this common  strategy used in the previous studies is suboptimal. 

For negative pairs, when $\eta$ is too small, we end the training process with a very small subset of the training data consisting of highly-ranked negative documents. In this case, the sampled documents may contain false negatives. The best performance is obtained with $\eta=0.7$, \ie, some mixture of easy and hard negative pairs is used at the end. However, when $\eta$ is too large (\ie, ~1.0), all negative pairs are randomly sampled during the whole training, the model's performance also decreases because of the disabled curriculum effect. These observations confirm the impact of both curricula and  suggest that the right degree of difficulty in the initial and final pools of samples may influence the effectiveness of \ttt{DCL}. 

\subsubsection{Performance on Sessions with Different Lengths}
To understand the impact of the session length on the final ranking performance, we categorize the sessions in the test set into three bins: 

(1) Short sessions (with 1-2 queries) - 77.13\% of the test set;

(2) Medium sessions (with 3-4 queries) - 18.19\% of the test set;

(3) Long sessions (with 5+ queries) - 4.69\% of the test set.

We compare \ttt{COCA+DCL} with \ttt{Duet}, \ttt{HBA}, \ttt{RICR}, and \ttt{COCA} and show the results regarding MAP and NDCG@3 in the left side of Figure~\ref{fig:len}. First, \ttt{COCA+DCL} improves the performance of \ttt{COCA} across all three session bins. This shows \ttt{DCL}'s high robustness for different kinds of search context. Second, we can see the ad-hoc ranking method \ttt{Duet} performs worse than other context-aware ranking methods. This highlights once again that modeling historical user behavior is essential for improving document ranking performance. Third, \ttt{COCA+DCL} performs better on short sessions than on long ones. We hypothesize that those longer sessions are inherently more difficult to understand, and a similar trend in baseline methods may corroborate this. This can be due to the fact that a long session may contain more diverse intents or exploratory search. 

\subsubsection{Effect of Modeling User Behavior Progression}
It is important to study how the modeled search context contributes to document ranking as a search session  progresses. We compare \ttt{COCA+DCL} with \ttt{HBA}, \ttt{RICR}, and \ttt{COCA} at individual query positions in short (S), medium (M), and long (L) sessions. The results are presented in the right side of Figure~\ref{fig:len}. Due to space limitations, long sessions with more than seven queries are omitted.

We can see that the ranking performance generally improves when the short and medium sessions progress (\eg, S2 is higher than S1) because more search context information becomes available for predicting the next click. It benefits \ttt{COCA+DCL} and two baselines (\ttt{COCA} and \ttt{HBA}), while \ttt{COCA+DCL} improves more rapidly by better exploiting the context. 
One interesting observation is that, 
when the search sessions become longer (\eg, from L4 to L7), the gain of \ttt{DCL} decreases. We attribute this to the noisier nature of long sessions.

\section{Conclusion and Future Work}
In this work, we proposed a novel curriculum learning framework for context-aware document ranking. Two complementary curricula were designed for learning positive and negative context-document pairs in an easy-to-hard manner. With these curricula, the model's capability of capturing matching signals and identifying mismatching signals is gradually enhanced. We conducted experiments with three recently proposed methods on two large-scale datasets. The results clearly demonstrate the effectiveness and wide applicability of our framework. Besides, we also investigated the influence of different settings on applying curriculum learning to context-aware document ranking.
Our work is an early attempt to apply curriculum learning to IR, and there is still much space to be explored. For example, it may be useful to mine the most valuable negative candidate documents. Considering query and document weighting in computing the difficulty is also a future direction.

\begin{acks}
Zhicheng Dou is the corresponding author. This work was supported by a Discovery grant of the Natural Science and Engineering Research Council of Canada, National Natural Science Foundation of China (No. 61872370), Beijing Outstanding Young Scientist Program (No. BJJWZYJH012019100020098), and Beijing Academy of Artificial Intelligence (BAAI).
\end{acks}

\clearpage
\balance
\bibliographystyle{ACM-Reference-Format}
\bibliography{sample-base}
\end{document}